\documentclass[conference]{IEEEtran}
\IEEEoverridecommandlockouts
\let\oldbibliography\thebibliography
\renewcommand{\thebibliography}[1]{%
  \oldbibliography{#1}%
  \setlength{\itemsep}{0pt}}   
  
\usepackage{amsmath,amssymb,amsfonts}
\usepackage{graphicx}
\usepackage{pgfplots}
\usepackage{tikz}
\usetikzlibrary{matrix, shapes.geometric, arrows}
\usepackage{multirow}
\usepackage{array}
\usepackage{booktabs}
\usepackage{cite}
\usepackage{algpseudocode}
\usepackage{textcomp}
\usepackage{xcolor}
\usepackage{url}
\usepackage{wrapfig}
\usepackage{balance}
\usepackage{subcaption} 
\usepackage[font=footnotesize,labelfont=bf]{caption}
\usepackage[none]{hyphenat} 

\begin{document}
\title{Identification of High Impedance Faults Utilizing Recurrence Plots} 
\author{

{\tt\small}
Pallav Kumar Bera\\
Western Kentucky University\\
KY, USA \\
{\tt\small pallav.bera@wku.edu\vspace*{-0.1 cm}}

\and{\tt\small}
Samita Rani Pani\\
KIIT Deemed to be University \\
Odisha, India \\
{\tt\small samitapani.fel@kiit.ac.in\vspace*{-0.1 cm}}

\and{\tt\small}
Rajesh Kumar\\
Bucknell University \\
PA, USA \\
{\tt\small rajesh.kumar@bucknell.edu\vspace*{-0.1 cm}}

\thanks{© 2024 IEEE. Personal use of this material is permitted. Permission from IEEE must be obtained for all other uses, in any current or future media, including reprinting/republishing this material for advertising or promotional purposes, creating new collective works, for resale or redistribution to servers or lists, or reuse of any copyrighted component of this work in other works.
}}

\maketitle
\begin{abstract}

This paper presents a systematic approach to detecting High Impedance Faults (HIFs) in medium voltage distribution networks using recurrence plots and machine learning. We first simulate 1150 internal faults, including 300 HIFs, 1000 external faults, and 40 normal conditions using the PSCAD/EMTDC software. Key features are extracted from the 3-phase differential currents using wavelet coefficients, which are then converted into recurrence matrices. A multi-stage classification framework is employed, where the first classification stage identifies internal faults, and the second stage distinguishes HIFs from other internal faults. The framework is evaluated using accuracy, precision, recall, and F1 score. Tree-based classifiers, particularly Random Forest and Decision Tree, achieve superior performance, with 99.24\% accuracy in the first stage and 98.26\% in the second. The results demonstrate the effectiveness of integrating recurrence analysis with machine learning for fault detection in power distribution networks.
\end{abstract}

\begin{IEEEkeywords}
 High Impedance Faults, Random Forest, Machine Learning, Wavelet coefficients, Recurrence plots
\end{IEEEkeywords}

\section{Introduction}
Recent advancements in sensor technology, communication systems, and renewable energy integration are accelerating the deployment of smart power distribution networks. These changes impact conventional protection techniques due to modifications in the traditional radial system structure, including injecting intermediate currents and bidirectional current flow during faults \cite{GHADERI2017376}. High Impedance Faults (HIFs), common in distribution systems, often have low magnitudes and resemble normal disturbances, allowing them to bypass traditional protection systems \cite{wei}. Therefore, it is essential to develop fault detection methods to reliably identify HIFs under various operational scenarios in microgrids and distribution networks.

HIFs occur when an energized conductor contacts the ground through a high-impedance object, such as dry asphalt, wet sand, dry sod, or dry grass, which limits current flow to the ground. Conventional over-current protection schemes often fail to detect these faults because the fault current is significantly lower than the nominal load current \cite{aucoin}. Type-1 internal faults, including line-to-ground ($lg$), line-to-line ($ll$), and three-phase faults ($lll$), are easier to detect due to the noticeable changes in system currents and voltages. In contrast, Type-2 faults, or HIFs, are difficult to detect because of their lower fault current levels, which can resemble normal load currents. These faults typically involve a downed conductor or contact with high-impedance surfaces, resulting in arcing, asymmetry, and intermittent currents.

HIF detection methods range from time and frequency domain algorithms to time-frequency analyses and AI techniques \cite{gao}. For example, \cite{ASGHARIGOVAR2019} combined wavelet transform (WT) with an extreme learning machine (ELM). Similarly, \cite{song} utilized the Hilbert-Huang Transform (HHT) with a negative selection algorithm. \cite{SHENG2024} extended this approach with a kernel ELM incorporating WT and HHT features. Real-time detection using a convolutional neural network (CNN) and short-time Fourier transform (STFT) was demonstrated by \cite{sirojan}. Additionally, \cite{yong} applied decision trees (DTs) based on phase current and harmonic analysis. \cite{guo} employed zero-sequence current analysis via continuous wavelet transform (CWT) combined with a CNN for detection.
Bokka et al. \cite{bokka} presented a novel technique using variational mode decomposition (VMD) for fault current feature extraction and HIF detection with a support vector machine (SVM). Complementary approaches include using Random Forest in \cite{pallavpv} and InceptionTime Networks in \cite{bera2023} to differentiate HIFs from other transient events.

While time-frequency analysis methods like WT and HHT effectively extract diverse features, a unified and systematic feature extraction and selection framework remains underdeveloped. The nonlinearity and stochastic nature of fault currents further challenge the reliability of traditional threshold-based detection methods.

This paper presents a methodology for extracting, selecting, and transforming features from 3-phase differential currents to develop a reliable fault detection system. Seven hundred eighty-three features are extracted, including statistical, frequency-domain, time-frequency, and entropy features. A Random Forest algorithm selected the most relevant features, reducing classification errors. The selected features were used to generate a recurrence matrix, input for machine learning algorithms. These algorithms are trained to detect internal faults and distinguish between Type-1 and Type-2 faults (HIFs). This approach eliminates the need for preset thresholds, enhancing detection accuracy.

The rest of the paper is organized as follows. Section~\ref{ModelingAndSimulation} details the test system, HIF model, and simulation of the transients, including the HIFs. Section~\ref{ClassificationFramework} explains the feature extraction, selection, and classification methods. Section~\ref{ResultsDiscussion} presents the classifier performance in fault detection and differentiation. Section~\ref{ConclusionFutureWork} concludes the study and suggests future research directions.

\begin{table}[h]
\centering
\renewcommand{\arraystretch}{1.2}
\setlength{\tabcolsep}{6pt}
\caption{System Components and their Specifications}
\label{mg_parameters}
\begin{tabular}{|l|l|}
\hline
\multicolumn{1}{|c|}{\textbf{Component}} & \multicolumn{1}{c|}{\textbf{Specification}} \\ \hline \hline
\textbf{Grid} & \begin{tabular}[c]{@{}l@{}}Frequency: 60 Hz \\ Voltage: 230 kV\end{tabular} \\ \hline
\multirow{3}{*}{\textbf{DG Sources}} & \begin{tabular}[c]{@{}l@{}}DG$'$ (Diesel): 2 MW, 13.8 kV, 0.833 kA\end{tabular} \\ \cline{2-2} 
 & \begin{tabular}[c]{@{}l@{}}DG$''$ (PV): 0.25 MW, 28\textdegree{}C, 1000 W/m\end{tabular} \\ \cline{2-2} 
 & \begin{tabular}[c]{@{}l@{}}DG$'''$ (Wind): 2.5 MW, 11 m/s\end{tabular} \\ \hline
\multirow{4}{*}{\textbf{Distribution Lines}} & DL21: 10 km \\ \cline{2-2} 
 & DL23: 20 km \\ \cline{2-2} 
 & DL24: 30 km \\ \cline{2-2} 
 & DL45: 10 km \\ \hline
\multirow{4}{*}{\textbf{Transformers}} & TX1: 230 kV/20 kV, 100 MVA \\ \cline{2-2} 
 & TX2: 13.86 kV/20 kV, 3 MVA \\ \cline{2-2} 
 & TX3: 460 V/20 kV, 1 MVA \\ \cline{2-2} 
 & TX4: 690 V/20 kV, 5.5 MVA \\ \hline
\multirow{4}{*}{\textbf{Loads}} & Load1: 0.95 MW, 0.475 MVAR \\ \cline{2-2} 
 & Load2: 0.95 MW, 0.475 MVAR \\ \cline{2-2} 
 & Load3: 0.05 MW, 0.025 MVAR \\ \cline{2-2} 
 & Load5: 0.95 MW, 0.475 MVAR \\ \hline
\end{tabular}
\end{table}

\begin{figure}[ht]
\centering
\includegraphics [width=3.4 in, height= 3.2 in] {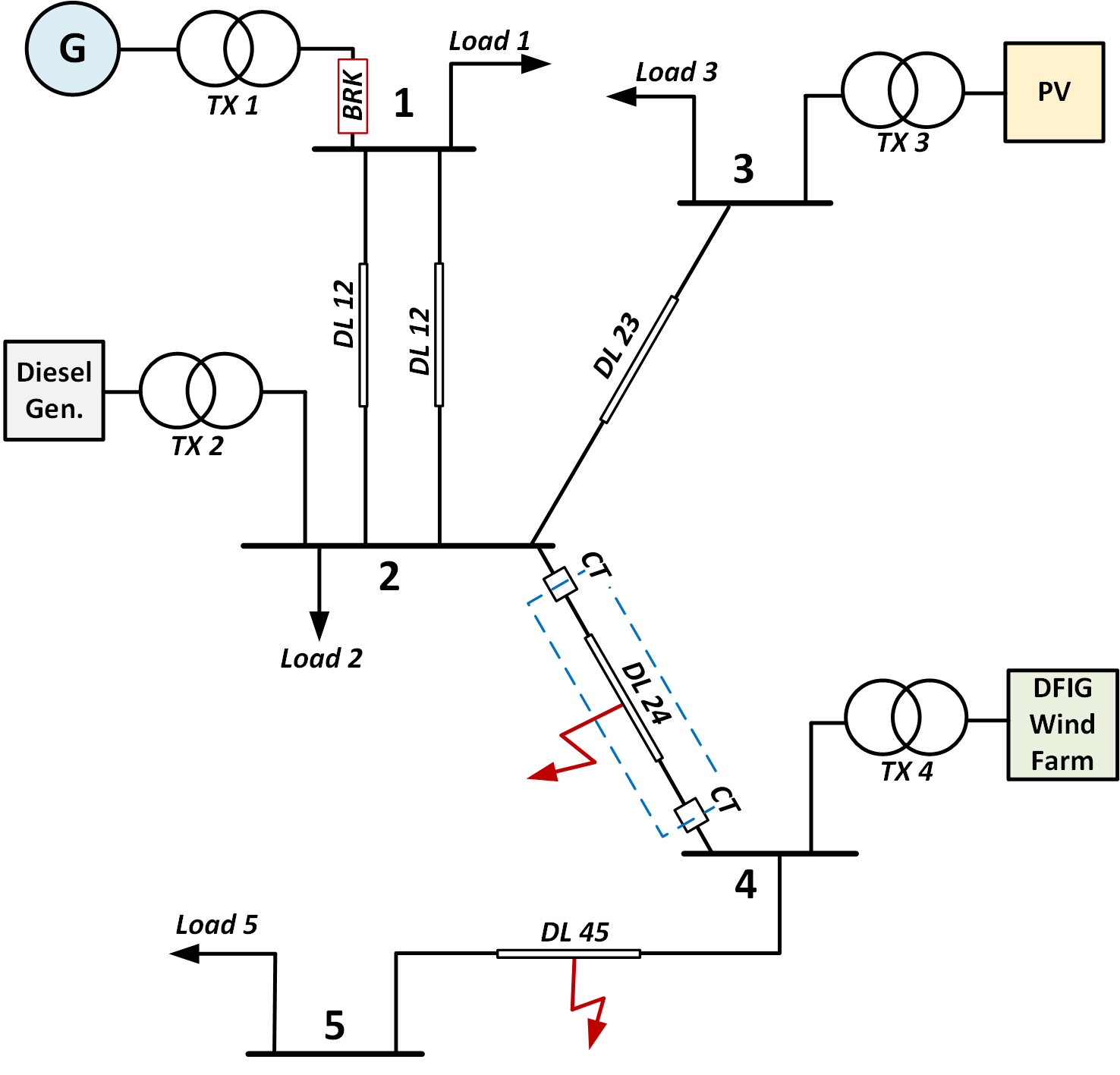} 
\caption{Test System with main grid, PV, Diesel Generator, and Wind Farm.}
\label{mg}
\end{figure}

\section{Modeling and Simulations}
\label{ModelingAndSimulation}
This section covers the development of the microgrid model using PSCAD/EMTDC, HIF modeling, and simulations for fault and normal scenarios.

\subsection{System Description}
The 5-bus microgrid with a diesel generator (DG$'$), photovoltaic (DG$''$), and wind farm (DG$'''$) is shown in Figure \ref{mg}. Component details are presented in Table \ref{mg_parameters}.

\subsection{High Impedance Fault Modeling}
\begin{figure}[ht]
\centering
\includegraphics[width=1.9in, height=1.4in]{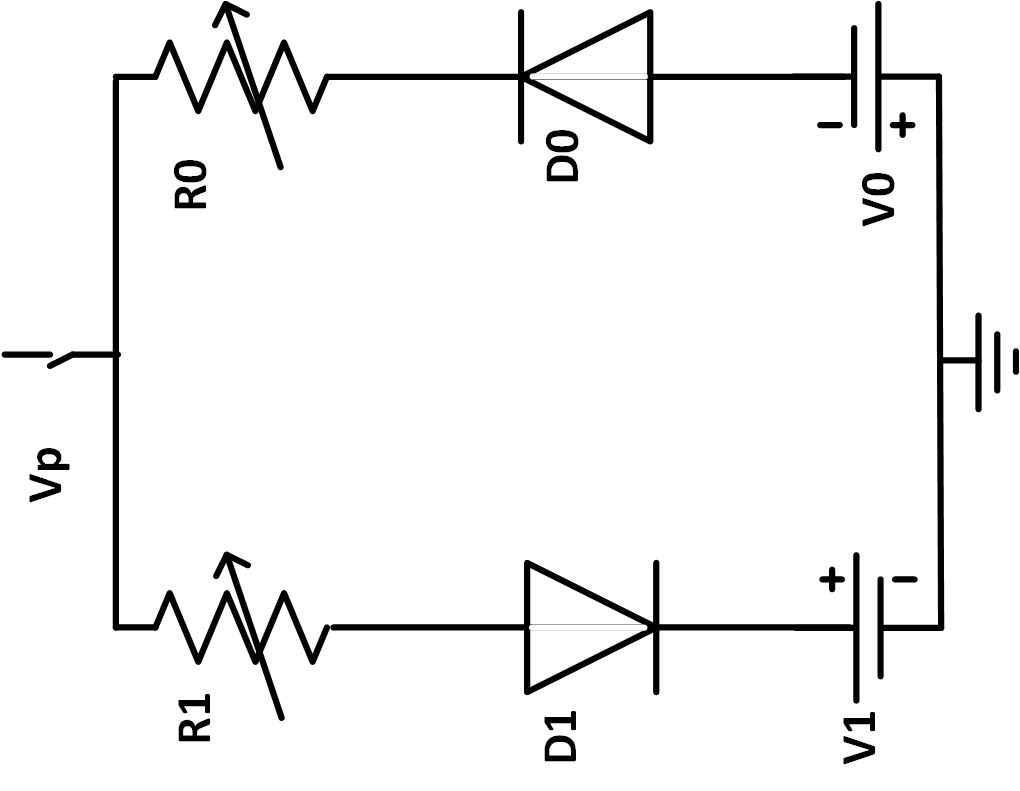}
\caption{Circuit diagram of the High Impedance Fault model.}
\label{hif}
\end{figure}

HIFs can be modeled in three ways \cite{2019}. This study uses the model with two anti-parallel DC sources, diodes, and variable resistors (Figure \ref{hif}) \cite{berahif}.
The variable resistors simulate the dynamic arc, and the changing DC sources represent fault current asymmetry. In the positive half-cycle, \( V_p > V_1 \); in the negative half-cycle, \( V_p < V_0 \); and current is zero when \( V_1 < V_p < V_0 \).

\subsection{Fault and Normal Data}
Internal faults, including HIFs, are simulated on the DL24 line between bus-2 and bus-4. Type 1 internal faults include line-to-ground (\(lg\)), line-to-line-to-ground (\(llg\)), line-to-line (\(ll\)), and various three-phase faults (\(lllg\), \(lll\)). In grid-connected mode, 400 faults are simulated by varying fault types (5 types), resistance (5 different resistance values), and fault inception time for balanced and unbalanced loads. In islanded mode, 450 faults are simulated by varying fault types (5), resistance (5), and fault inception time (6 times) for balanced loads, unbalanced loads, and low voltage conditions. Thus, a total of 850 internal faults are obtained.

\begin{figure}[ht]
\centering
\includegraphics[width=3.3in, height=3.1in]{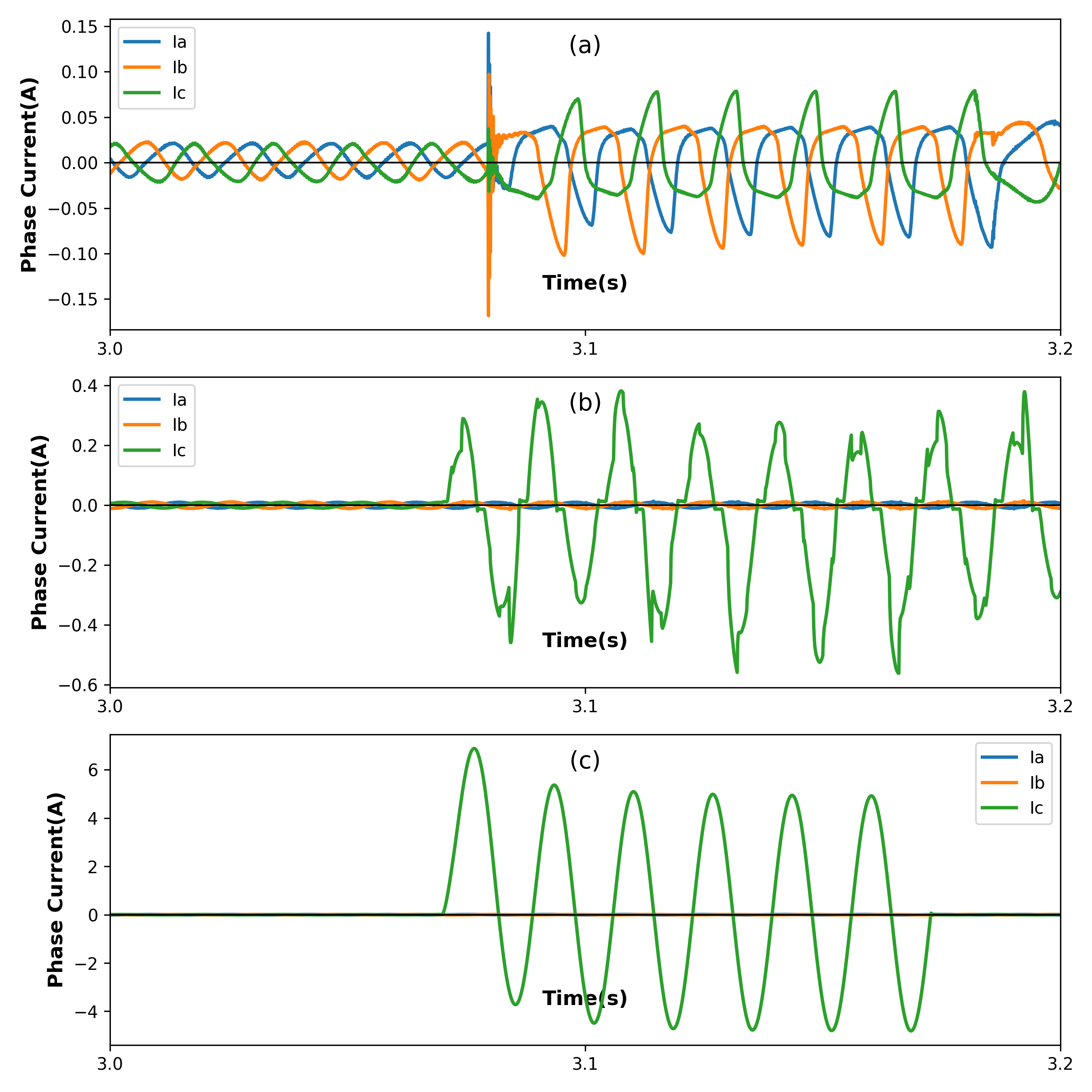} 
\caption{Differential currents showing (a) external faults with CT saturation, (b) High Impedance Fault, and (c) internal faults across phases a, b, and c.}
\label{faults}
\end{figure}

HIFs, or type 2 internal faults, are often associated with downed conductors. Fault impedance randomly varies between 50 $\Omega$ and 300 $\Omega$ within a 0.2 ms interval. Voltages as low as 0.9 per unit are also considered. 
In grid-connected mode, 120 HIFs are simulated by varying the phase (3 phases) and fault inception time (20) for balanced and unbalanced loads. In islanded mode, 180 HIFs are simulated by varying the phase (3) and fault inception time (20) for balanced, unbalanced loads and low voltage conditions. Thus, a total of 300 HIFs are obtained.

External faults are simulated on line DL45, with CTs on line DL24 subjected to different burdens. In grid-connected mode, 400 external faults are simulated by varying fault type (5), resistance (4), and fault inception time (10) for balanced and unbalanced loads. Similarly, 600 faults are simulated in islanded mode by varying the same parameters for balanced, unbalanced loads and low voltage conditions. Thus, a total of 1000 external faults with CT saturation are obtained.

Additionally, 40 normal conditions are simulated by varying the line length of DL24. The 3-phase differential currents for internal faults, HIFs, and external faults with CT saturation are shown in Figure \ref{faults}.

\section{Classification Framework}
\label{ClassificationFramework}
This section outlines the classification framework, covering feature extraction, feature selection, identification of the most significant feature, computation of the Wavelet Coefficient, and transformation into a Recurrence Matrix.

\subsection{Multi-Stage Classification Framework}
\begin{figure}[ht]
\centering
\includegraphics[width=2.5in, height=2.8in]{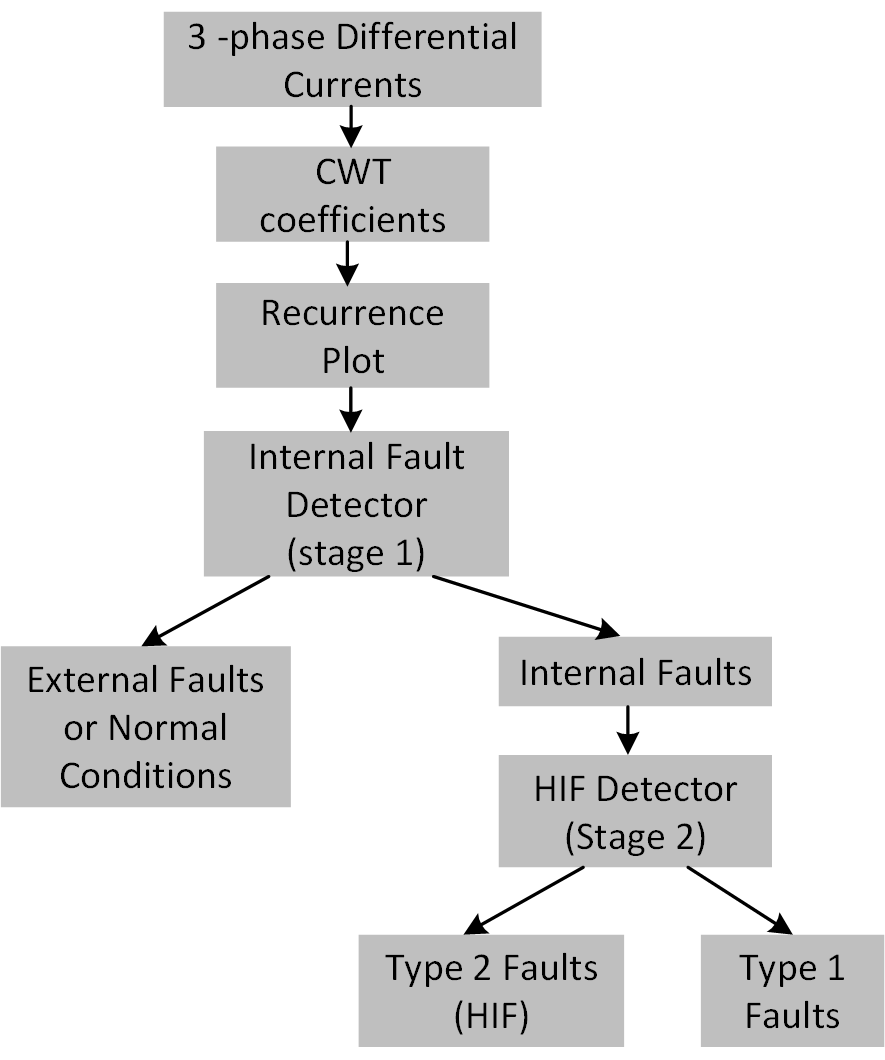}\caption{The proposed multi-stage classification framework. Stage 1 focuses on internal fault detection, while Stage 2 targets the identification of High Impedance Faults.}
\label{ms}
\end{figure}

The classification problem is tackled through a multi-stage framework, dividing the task into two distinct stages, as depicted in Figure \ref{ms}.

\subsection{Feature Extraction}
Feature extraction involves turning raw data into relevant characteristics to improve machine learning model accuracy. A comprehensive set of 783 features, including statistical (mean, variance), frequency (FFT coefficients), time-frequency (wavelet coefficients), entropy (sample entropy, approximate entropy), and temporal (autocorrelation, number of peaks) features, is evaluated in Python. A complete list is provided in \cite{tsfresh}. The best features from the 3-phase differential currents measured by the CTs are used to train the classifier models.

Feature selection identifies the most important characteristics within differential currents, focusing on informative and independent features for accurate pattern recognition and classification. This process reduces input size and enhances reliability. Studies like \cite{pallavsystem} have used time, spectral, and time-spectral domain features to classify power system transients. Features with higher information gain are more effective in distinguishing between fault and non-fault events. Feature selection reduces dimensionality, leading to faster computation and a more scalable, robust model.

\subsection{Random Forest Feature Selection}
Random Forest, an ensemble learning technique, performs implicit feature selection by evaluating how much each feature reduces impurity \cite{leo}. It trains multiple decision trees on random subsets of data, selecting a random subset of features at each node. The significance of a feature is determined by the reduction in entropy before and after the split. Entropy \( H(F) \) for a feature \( F \) is calculated as:

\[
H(F) = - \sum_{i=1}^{n} P(f_i) \log_2 P(f_i)
\]

where \( P(f_i) \) is the probability of the \( i \)-th class in \( F \).

Information gain \( IG(D, F) \), the reduction in entropy after splitting dataset \( D \) by feature \( F \), is given by:

\[
IG(D, F) = H(Y) - \sum_{v \in \text{Values}(F)} \frac{|D_v|}{|D|} H(D_v)
\]

Feature importance \( FI(F) \) in Random Forest is calculated by averaging entropy reduction across all trees:

\[
FI(F) = \frac{1}{T} \sum_{t=1}^{T} \sum_{n \in \text{Nodes}(t, F)} \Delta H_n
\]

This process ranks features based on their contribution to reducing uncertainty. 
The average entropy reduction ranks the features with higher values signifying more critical features. It enhances the robustness of feature selection by considering various subsets and combinations and helps
identify the most influential features that contribute to
accurate model predictions. By leveraging the ensemble nature
of Random Forest, the proposed method ensures that the selected
features are consistently valuable across different scenarios,
leading to more reliable models.

Through this method, Random Forest identified the most significant features, ultimately leading to the selection of Continuous Wavelet coefficients with a scale of ten and a shift of three.

\subsection{Wavelet Coefficients}
The Ricker wavelet, $\phi(t, g, h)$, is obtained by scaling and shifting the mother wavelet $\phi(t)$, expressed as:

\begin{equation}\label{mhw}
\phi(t,g,h) = \frac{2}{\sqrt{3g} \pi^{\frac{1}{4}}} \left(1 - \frac{(t-h)^2}{g^2}\right) \exp\left(-\frac{(t-h)^2}{2g^2}\right)
\end{equation}

Here, $g$ and $h$ are the scaling and shifting parameters. The Continuous Wavelet Transform (CWT) of a signal $y(t)$ is given by:

\begin{equation}\label{cwt}
Y(y(t),g,h,\phi(t)) = \int_{-\infty}^{+\infty} y(t) \phi^*(t,g,h) \, dt
\end{equation}

where $\phi^*(t,g,h)$ is the complex conjugate of the wavelet function.

\begin{figure}[ht]
\centering
\includegraphics[width=3.5in, height=2.4in]{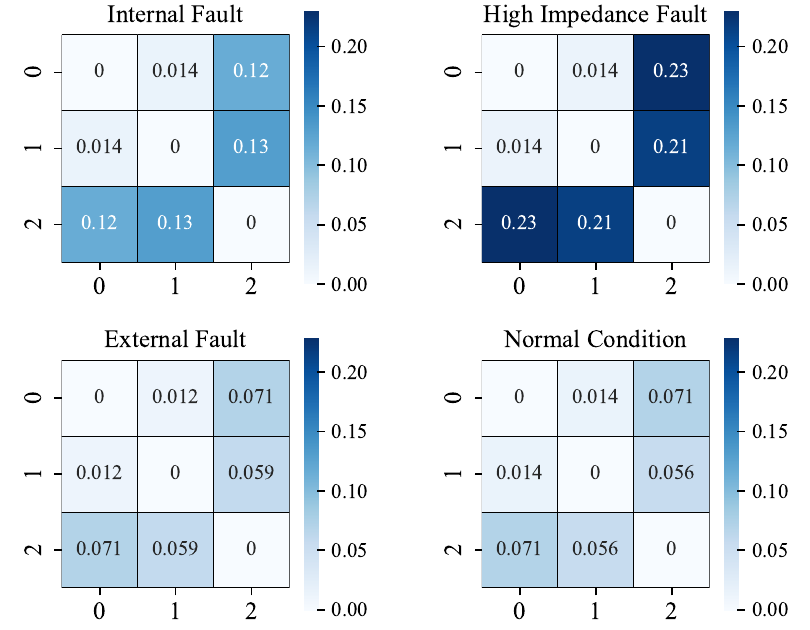} 
\caption{Recurrence matrix heatmaps (recurrence plots) illustrating the distinction between (a) internal fault, (b) external faults with CT saturation, (c) High Impedance Fault, and (d) normal operating conditions.}
\label{rp}
\end{figure}

\subsection{Recurrence matrix}
The recurrence plot, introduced by Eckmann \cite{rp}, visualizes the recurrence of states in a time series. Traditionally, it uses a binary matrix to represent these recurrences. In this study, the recurrence matrix is constructed using Euclidean distances between states for a deeper analysis of time series relationships.

For a time series \(\{s_i\}_{i=1}^N\), the recurrence plot is generated by comparing all pairs of points in the phase space trajectory. This trajectory is derived by embedding the time series into a higher-dimensional space using delay embedding.

The recurrence matrix \(\mathbf{R}\) is defined as an \(N \times N\) matrix, where each element \(R_{i,j}\) represents the Euclidean distance between states \(s_i\) and \(s_j\):

\[
R_{i,j} = \| \mathbf{s}_i - \mathbf{s}_j \|,
\]

Where:
\begin{itemize}
    \item \(\mathbf{s}_i = (s_i, s_{i+\tau}, \ldots, s_{i+(m-1)\tau})\) is the embedded vector,
    \item \(\tau\) is the time delay,
    \item \(m\) is the embedding dimension,
    \item \(\| \cdot \|\) is the Euclidean norm.
\end{itemize}

The recurrence matrix reveals patterns in dynamical behavior. Diagonal lines indicate deterministic or periodic behavior. Vertical and horizontal lines suggest steady states. Isolated points may indicate rare events or noise.

To compute the recurrence matrix from a time series \(\{s_i\}_{i=1}^N\), select the embedding dimension \(m\) and time delay \(\tau\), construct vectors \(\mathbf{s}_i\) for \(i = 1, 2, \ldots, N - (m-1)\tau\), and calculate the pairwise distances \(\|\mathbf{s}_i - \mathbf{s}_j\|\).

Given a time series \(\mathbf{s} = [-0.024, -0.008, 0.001]\), the pairwise Euclidean distances are:

1. Distance between \(s_1 = -0.024\) and \(s_2 = -0.008\):
\[
d_{12} = |s_1 - s_2| = 0.016
\]

2. Distance between \(s_1 = -0.024\) and \(s_3 = 0.001\):
\[
d_{13} = |s_1 - s_3| = 0.025
\]

3. Distance between \(s_2 = -0.008\) and \(s_3 = 0.001\):
\[
d_{23} = |s_2 - s_3| = 0.009
\]

The resulting distance matrix \(\mathbf{D}\) is:
\[
\mathbf{D} =
\begin{bmatrix}
d_{11} & d_{12} & d_{13} \\
d_{21} & d_{22} & d_{23} \\
d_{31} & d_{32} & d_{33}
\end{bmatrix} \]

\[=
\begin{bmatrix}
0 & 0.016 & 0.025 \\
0.016 & 0 & 0.009 \\
0.025 & 0.009 & 0
\end{bmatrix}
\]

The matrix \(D\) represents the pairwise Euclidean distances in the time series. The heatmap of the recurrence matrices like this for an internal fault, external fault with CT saturation, HIFs, and normal operation is shown in Figure \ref{rp}.

\begin{table*}[ht]
    \centering
    \caption{Performance Metrics of Various Classifiers for Internal Faults (Stage 1)}
    \begin{tabular}{lccccccc}
        \hline
        Classifier & Accuracy ($\eta_{1}$) & Precision & Recall & F1-Score & ROC-AUC & Train Time (s) & Test Time (s) \\
        \hline
        RF         & \textbf{0.9924} & 0.9924 & 0.9924 & 0.9924 & 0.9994 & 2.1301 & 0.0733 \\
        k-NN                  & 0.9848 & 0.9850 & 0.9848 & 0.9848 & 0.9968 & 0.0030 & 0.0212 \\
        DT        & \textbf{0.9924} & 0.9924 & 0.9924 & 0.9924 & 0.9924 & 0.0071 & 0.0000 \\
        GB   & \textbf{0.9924} & 0.9924 & 0.9924 & 0.9924 & 0.9989 & 1.5248 & 0.0070 \\
        AB             & 0.9833 & 0.9833 & 0.9833 & 0.9833 & 0.9944 & 1.2097 & 0.06869 \\
        MLP                   & 0.9513 & 0.9513 & 0.9513 & 0.9513 & 0.9524 & 1.2381 & 0.1303 \\
        \hline
    \end{tabular}
    \label{classifier_comparison1}
\end{table*}

\begin{table}[ht]
\centering
\caption{Comparison of Various Classifiers for HIF Detection (Stage 2)}
\begin{tabular}{lcccccc}
\toprule
Classifier & Accuracy ($\eta_{2}$) & Precision & Recall & F1-Score & AUC \\
\midrule
RF & 0.9797 & 0.9803 & 0.9797 & 0.9794 & 0.9997 \\
k-NN & 0.9652 & 0.9661 & 0.9652 & 0.9645 & 0.9703 \\
DT & \textbf{0.9826} & 0.9830 & 0.9826 & 0.9824 & 0.9667 \\
GB & 0.9681 & 0.9694 & 0.9681 & 0.9674 & 0.9992 \\
AB & \textbf{0.9826} & 0.9830 & 0.9826 & 0.9824 & 0.9798 \\
MLP & 0.7188 & 0.6867 & 0.7188 & 0.6958 & 0.8017 \\
\bottomrule
\end{tabular}
\label{classifier_comparison2}
\end{table}

\section{Results and Discussion}
\label{ResultsDiscussion}
This section presents the analysis of the fault detection performance based on the experimental dataset. The evaluation was divided into two key parts: the performance metrics used to assess the classifiers and the comparative analysis of different classification algorithms in detecting internal faults and HIFs.
 
\subsection{Performance Metrics}
The performance of the fault classification algorithms was measured using key metrics that quantified the accuracy and reliability of fault detection:

\begin{itemize}
    \item \textbf{Accuracy}: Measures the proportion of correctly classified instances, including fault and non-fault cases.
        \[
        \text{Accuracy} = \frac{N_{\text{TP}} + N_{\text{TN}}}{N_{\text{TP}} + N_{\text{TN}} + N_{\text{FP}} + N_{\text{FN}}}
        \]

    \item \textbf{Precision}: Evaluates the reliability of fault predictions by assessing how many predicted faults are actual faults.
        \[
        \text{Precision} = \frac{N_{\text{TP}}}{N_{\text{TP}} + N_{\text{FP}}}
        \]

    \item \textbf{Recall}: Measures the model's ability to correctly identify actual faults, reflecting the proportion of actual faults detected.
        \[
        \text{Recall} = \frac{N_{\text{TP}}}{N_{\text{TP}} + N_{\text{FN}}}
        \]

    \item \textbf{F1 Score}: Provides a balanced measure by combining Precision and Recall, which is especially useful for imbalanced datasets.
        \[
        \text{F1} = \frac{2 \times \text{Precision} \times \text{Recall}}{\text{Precision} + \text{Recall}}
        \]

    \item \textbf{ROC-AUC}: Evaluates the model's ability to distinguish between fault and non-fault classes.
        It measures the area under the curve plotting True Positive Rate (TPR) against False Positive Rate (FPR).
        \[
        \text{AUC} = \int_{0}^{1} \text{TPR}(\theta) \, d\text{FPR}(\theta)
        \]
 
\end{itemize}

\subsection{Performance of Classifiers}
Fault detection performance was evaluated using six classifiers, categorized into tree-based and non-tree-based methods. The tree-based classifiers were Random Forest (RF), Decision Tree (DT), Gradient Boosting (GB), and AdaBoost (AB). The non-tree-based methods were k-Nearest Neighbors (k-NN) and Multi-Layer Perceptron (MLP). The classifiers were tested on two tasks: distinguishing internal faults from external faults and normal conditions (Stage 1) and identifying HIFs from internal faults (Stage 2).

As shown in Table \ref{classifier_comparison1}, in Stage 1, the tree-based classifiers RF, DT, and GB achieved the highest accuracy of 99.24\%, with strong Precision, Recall, and F1 scores. These models effectively handled non-linear relationships and outliers. The k-NN classifier showed 98.48\% accuracy, slightly lower than the tree-based methods. AdaBoost recorded 98.33\% accuracy, while MLP had the lowest accuracy at 95.13\%.
RF and DT classifiers achieved near-perfect AUC values, highlighting their effectiveness in distinguishing internal faults from other conditions. Both classifiers had minimal training and testing times, making them efficient for real-time applications due to their balance between accuracy and computational efficiency. 

In Stage 2, focused on detecting HIFs, DT and AB classifiers achieved the highest accuracy of 98.26\%, followed by RF at 97.97\% and GB at 96.81\% (Table \ref{classifier_comparison2}). The F1 scores and AUC values for these classifiers remained consistently high, indicating their reliability in identifying HIFs, which are challenging due to their subtle characteristics.
The k-NN classifier performed well in Stage 2, achieving an accuracy of 96.52\%, confirming its utility in fault detection. However, MLP performed poorly, achieving only 71.88\% accuracy, indicating its limitations in HIF detection compared to tree-based methods.

Tree-based classifiers (RF, DT, GB, and AB) consistently outperformed non-tree-based models in both stages. These classifiers effectively modeled complex, non-linear relationships and handled imbalanced datasets without overfitting. While k-NN performed well, it did not match the Precision and Recall of the top tree-based models.
MLP performed poorly in this context, likely because its neural network structure required further tuning for fault data. In contrast, tree-based models are easier to implement and less prone to overfitting.

RF, DT, GB, and AB proved to be the most effective for internal fault detection and HIF identification. Their high accuracy, Precision, Recall, and strong AUC scores make them suitable for fault detection in distribution networks. These results emphasize the importance of ensemble learning and tree-based models in handling complex fault detection tasks with reliability and efficiency.

\section{Conclusion and Future Work}
\label{ConclusionFutureWork}

This study demonstrated the efficacy of a machine learning-based differential protection strategy for medium-voltage distribution lines in a microgrid powered by renewable energy. By utilizing wavelet coefficients derived from 3-phase current measurements and converting them into recurrence matrices, the approach achieved 99.24\% accuracy in identifying internal faults, including low-current High Impedance Faults, from external faults and normal conditions. Tree-based classifiers, particularly Random Forest and Decision Trees, proved the most effective, highlighting their capability to handle complex, non-linear relationships and accurately distinguish between fault types.

Future work could explore applying this approach in various electrical networks, including those with different configurations and operational conditions. Additionally, refining the models to accommodate a broader range of fault scenarios will enhance the adaptability and robustness of the proposed protection strategy, further contributing to reliable and efficient fault detection in diverse power systems.

\balance
\bibliography{ref} 
\bibliographystyle{ieeetr}
\end{document}